\documentclass[sigconf,nonacm=true]{aamas}

\setcopyright{none}

\usepackage{geometry}
\usepackage{times}
\usepackage[disable,colorinlistoftodos]{todonotes}
\usepackage{fancyvrb}
\usepackage{url}
\usepackage{comment}

\renewcommand\footnotetextcopyrightpermission[1]{}
\newgeometry{textwidth=177.9mm, textheight=222mm, columnsep=2pc}

\begin{document}

\title{Incorporating social practices in BDI agent systems}  

\author{Stephen Cranefield}
\affiliation{%
  \institution{University of Otago}
  \city{Dunedin}
  \state{New Zealand}
  \postcode{9010}
}
\email{stephen.cranefield@otago.ac.nz}
\author{Frank Dignum}
\affiliation{%
  \institution{Ume\aa{} University}
  \city{Ume\aa}
  \state{Sweden}
}
\email{frank.dignum@umu.se}

\begin{abstract}
  When agents interact with humans, either through embodied agents or because they are embedded in a robot, it would be easy if they could use fixed interaction protocols as they do with other agents. However, people do not keep fixed protocols in their day-to-day interactions and the environments are often dynamic, making it impossible to use fixed protocols. Deliberating about interactions from fundamentals is not very scalable either, because in that case all possible reactions of a user have to be considered in the plans. In this paper we argue that social practices can be used as an inspiration for designing flexible and scalable interaction mechanisms that are also robust. However, using social practices requires extending the traditional BDI deliberation cycle to monitor landmark states and perform expected actions by leveraging existing plans. We define and implement this mechanism in Jason using a periodically run meta-deliberation plan, supported by a metainterpreter, and illustrate its use in a realistic scenario.
\end{abstract}

\keywords{Social practices; BDI agents; Jason}  

\maketitle


\section{Introduction}
\label{sec:introduction}

Imagine the scenario where a disabled person, living alone, is assisted by a care robot. The robot takes care that the person gets up every morning and makes sure that he drinks some coffee and takes his morning pills (if needed). Then they read the newspaper, which means that the person looks at the pictures in the paper and the robot reads the articles out loud for the person to hear. (The hearing of the person is better than his eyesight, so, he cannot read the small font of the newspaper well, but can hear the robot).

When agents in the role of this type of personal assistant or care robot have to interact with humans over a longer time period and in a dynamic environment (that is not controlled by the agent), the interaction management becomes very difficult. When fixed protocols are used for the interaction they are often not appropriate in all situations and cause breakdowns and consequent loss of trust in the system. However, to have real-time deliberation about the best response during the interaction is not very scalable, because in real life the contexts are dynamic and complex and thus the agent would need to take many parameters into consideration at each step. Thus we need something in between a completely scripted interaction that is too brittle and a completely open interaction that is not scalable.

As we have done before in the agent community, we take inspiration from human interactions and the way they are managed by individuals. We classify situations into standard contexts in which a certain \emph{social practice} can be applied. Social science has studied this phenomenon in social practice theory. Social practice theory comes forth from a variety of different sub-disciplines of social science. It started from philosophical sociology with proponents like Bourdieu \cite{Bourdieu1976} and Giddens \cite{Giddens1979}. Later on Reckwitz and Shove \cite{Reckwitz,shove2012} have expanded on these ideas, and also Schatzki \cite{Schatzki2012} made some valuable contributions.

These authors all claim that important features of human life should be understood in terms of organized constellations of interacting persons. Thus people are not just creating these practices, but our deliberations are also based on the fact that most of our life is shaped by social practices. Thus we use social practices to categorize situations and decide upon ways of behaviour based on social practices. The main intuition behind this is that our life is quite cyclic, in that many activities come back with a certain regularity. We have meals every day, go to work on Monday until Friday, go to the supermarket once a week, etc. These so-called Patterns of Life (\cite{Flosom2014}) can be exploited to create standard situations and expectations. It makes sense to categorize recurrent situations as social practices with a kind of standard behaviour for each of them.

Unfortunately social practice theory has not been widely used in computer science or in HCI and thus there are no ready-to-use tools in order to incorporate them in agents. It is clear from the above description that social practices are more than just a protocol or a frame to be used by the agent in its deliberation. Therefore, in this paper we make the following contributions. We propose a mechanism for BDI agents to maintain awareness about active social practices, and to leverage their existing plans to act in accordance with these practices. This is presented as a meta-deliberation plan that can be directly executed by Jason agents, or treated as a specification for an optimised implementation in an extended agent platform. This plan has been deployed in the (simulated) care robot scenario, to confirm that awareness of and adherence to a social practice enables the robot to have a more successful interaction with the patient over a longer period of time. As some of the features needed to implement this scenario, and to support our meta-deliberation plan, are not currently available in Jason, we also present a Jason metainterpreter, which provides this extended functionality, but can also be used independently to support other research on extensions to practical reasoning in the BDI agent paradigm.

In the next section, we give an introduction to the purpose and structure of social practices. In Section~\ref{sec:scenario}, we elaborate on the care robot scenario and how we have modelled it in Jason. In Section~\ref{sec:scenario_revisited}, we describe the role of social practices in this scenario, and discuss the requirements this imposes for a BDI agent. In Section \ref{sec:implementation}, we present our mechanism for extending Jason to leverage social practices, and the metainterpreter needed to support this. We finish the paper with some conclusions and suggestions for future work.

\enlargethispage{0.5\baselineskip}
\section{Social Practices}
\label{sec:SPs}

Social practices are defined as accepted ways of doing things, contextual and materially mediated, that are shared between actors and routinized over time \cite{Reckwitz}. They can be seen as patterns which can be filled in by a multitude of single and often unique actions. Through (joint) performance, the patterns provided by the practice are filled out and reproduced.

According to \cite{Reckwitz,shove2012} a social practice consists of three parts: 
\begin{itemize}
\item Material: covers all physical aspects of the performance of a practice, including the human body and objects that are available (relates to physical aspects of a context).
\item Meaning: refers to the issues which are considered to be relevant with respect to that material, i.e. understandings, beliefs and emotions (relates to social aspects of a situation).
\item Competence:  refers to skills and knowledge which are required to perform the practice (relates to the notion of deliberation about a situation).
\end{itemize}

Whereas the first and third parts intuitively can be made more precise for an implementation, the second part is rather vague. Let us consider these three parts of a social practice in the scenario of the care robot scenario introduced in Section~\ref{sec:introduction}.
The material refers to the room where the robot serves morning coffee for the disabled person. It includes the materials that are needed to make coffee (such as coffee and a coffee maker) and serve it (such as a cup and tray). However, it also includes the table and other furniture in the room, the newspaper (if present), the TV, radio, computer, tablet, and the robot and person (and possible other people that are present).

The competence part describes what activities every party can perform and expectations about what they will actually do. For example, the robot is capable of making coffee and serving it. The person can drink his coffee by himself. They both can read the newspaper or watch TV. The expectation is that the robot wakes the person if he is not awake yet, makes the coffee and gives it to the person. After that they will read the newspaper together to provide mental stimulation. Note, these are expectations, not a protocol. So, parties can deviate from it and they can also fill the parts in, in ways they see fit best.
The meaning part has to do with all the social interpretations that come with the social practice, e.g.~drinking coffee in the morning
might give the person a sense of well-being that he can use to face the challenges of the rest of the day. When the coffee is cold or weak the person might interpret it as disinterest on the part of the robot in his well-being. The goal of reading the newspaper might also be not just to get the information from it, but a form of entertainment and feeling related to the robot, because you do something together.

Given the above descriptions of social practices one can summarize the purpose of them for individuals (informally) as follows:
\[ 
Perceive(resources) \Rightarrow Expect(activities) \wedge Expect(competencies)
\]
\[
Done(activities) \Rightarrow physical(postconditions) \wedge social(postconditions)
\]
Given a certain situation that is perceived in which the social practice is activated, the social practice now triggers expectations that the activities of that social practice will be executed. This implies that we assume a certain competence of all the people involved in the social practice (e.g. the robot can make coffee). After the activities have been executed, we do not just assume the postconditions of the actions hold, but also assume that certain social effects have been achieved according to the ``meaning'' of the practice. Thus the social practice allows us to make a whole set of assumptions and have expectations that could otherwise not be readily made or would take a lot of effort to derive.

From the above description it can already be seen that social practices are more encompassing than conventions and norms. Conventions focus on the strategic advantage that an individual gets by conforming to the convention. The reason to follow a convention is that if all parties involved comply, a kind of optimal coordination is reached, i.e.~if we all drive on the left side of the road, traffic will be smoother than when everyone chooses the side to drive on freely. Thus, conventions focus on the actual actions being performed and how they optimize the coordination. Social practices do not necessarily optimize the coordination. Because they indicate
expected actions and interactions given a social and physical context they will smoothen the coordination. However, this is not necessarily the optimal way the coordination could have been done. For example, if we go to a presentation, we sit down as soon as we see chairs standing in rows in the room. However, we could also keep standing (as is often done outside).

Social practices are also different from norms. Norms are also applicable in certain situations and for particular people (or roles) and they also create expectations (namely that the norm is followed). However, norms usually dictate a very specific behaviour rather than creating a set of loosely couple expectations as is the case for social practices. E.g. if the norm states that a car has to stop for a red light, it gives a very specific directive. If a norm is more abstract (like ``drive carefully'') then we need to translate this into concrete norms for specific situations. This is different from saying that some parts of a situation are governed by the norm and others are still free of the normative influence. Basically, when specifying the norm one indicates exactly when the norm is applicable rather than general situations for which the norm can be applied in some part.

One framework that seems very close to social practices is the notion of \emph{scripts}. However, social practices are not just mere scripts in the sense of Minsky \cite{Minsky}. Practices are more flexible than the classical frames defined by scripts in that they can be extended and changed by learning, and the ``slots'' only need to be filled in as far as they are needed to determine a course of action. Using these structures changes planning in many common situations to pattern recognition and filling in parameters. They support, rather than restrict, deliberation about behaviour. For example, the social practice of ``going to work'' incorporates usual means of transport that can be used, timing constraints, weather and traffic conditions, etc. So, normally you take a car to work, but if the weather is exceptionally bad, the social practice does not force the default action, but rather gives input for deliberation about a new plan in this situation, such as taking a bus or train (or even staying home). So, social practices can be seen as a kind of flexible script. Moreover, scripts do not incorporate any social meaning for the activities performed in them as social practices do.

Social practices have been used in applications already in a variety of ways. In \cite{Gilbert,Mercuur} they have been used as part of social simulations. In those applications, social practices are used as a standard package of actions with a special status. Thus individuals can use them with a certain probability given the circumstances are right. However, these applications do not use the internal structure of social practices for the planning of the individuals. Social practices have been used for applications in natural language and dialogue management in \cite{Gentile,Harel}. Here, the social practices are used to guide the planning process, but are geared towards a particular dialogue rather than as part of a more general interaction. 
In \cite{Miller} it is shown how social practices can be used by a traditional epistemic multi-agent planner to provide efficient and robust plans in cooperative settings. However, in this case the planner was not part of a BDI agent with its own goals and plans, but completely dedicated to finding a plan for the situation at hand. In \cite{COIN2014} a first structure of social practices was presented that is more amenable for the use by agents. The paper is, unfortunately, only conceptual and no implementation was made yet. In this paper we will follow the structure described in \cite{COIN2014}, but use only those parts that are relevant for our scenario. A complete account would be far too large to fit in the space available here, but we take those parts that seem to be core to the idea of social practices and show how they work with BDI agents in the Jason platform.

The complete structure for social practices (based on \cite{COIN2014}) is as follows:\\
\textit{\textbf{Context}}
    \begin{itemize}
    \item \textit{Roles} describe the competencies and expectations about a certain type of actor. Thus the robot is expected to be able to make a cup of coffee.
    \item \textit{Actors} are all people and autonomous systems involved, that have capability to reason and (inter)act. This indicates the agents that are expected to fulfil a part in the practice. In our scenario, these are the robot and the person.
    \item \textit{Resources} are objects that are used by the actions in the practice, such as cups, coffee, trays, curtains, and chairs. So, they are assumed to be available both for standard actions and for the planning within the practice.
    \item \textit{Affordances} are the properties of the context that permit social actions and depend on the match between context conditions and actor characteristics. For example, the bed might be used as a chair, or a mug as a cup.
    \item \textit{Places} indicates where all objects and actors are usually located relatively to each other, in space or time: the cups are in the cupboard in the kitchen, the person is in the chair (or in bed), etc.
    \end{itemize}
\textit{\textbf{Meaning}}
    \begin{itemize}
    \item \textit{Purpose} determines the social interpretation of actions and of certain physical situations. For example, the purpose of reading the newspaper is to get information about current affairs and to entertain the person.
    \item \textit{Promotes} indicates the values that are promoted (or demoted, by promoting the opposite) by the social practice. Giving coffee to the person will promote the value of ``caring''. 
    \item \textit{Counts-as} are rules of the type ``X counts as Y in C'' linking brute facts (X) and institutional facts (Y) in the context (C). E.g., reading the newspaper with the person counts as entertaining the person.
    \end{itemize}
\textit{\textbf{Expectations}}
    \begin{itemize}
    \item \textit{Plan patterns} describe usual patterns of actions defined by the \emph{landmarks} that are expected to occur (states of affairs around which the inter-agent coordination is structured). For example, the care robot first checks if the person is awake then makes sure there is coffee served.
    \item \textit{Norms} describe the rules of (expected) behaviour within the practice. E.g., the robot should ask the person if he wants coffee, before starting to make it.
    \item \textit{Strategies} indicate condition-action pairs that can occur at any time during the practice. E.g. if the person drops the coffee, the robot will clean it up. If the robot notices the person is asleep (again) it will try to wake him.
    \item A \textit{Start condition}, or trigger, indicates how the social practice starts, e.g., The practice of having morning coffee starts at 8$\,$am.
    \item A \textit{Duration}, or \textit{End condition}, indicates how the social practice ends, e.g., the morning routine takes around 45 minutes and ends when the newspaper is read and the coffee is finished.
    \end{itemize}
\textbf{\textit{Activities}}
    \begin{itemize}
    \item \textit{Possible actions} describes the expected actions by actors in the social practice, e.g. making coffee, reading the newspaper, and opening curtains.
    \item \textit{Requirements} indicate the type of capabilities or competences that the agent is expected to have in order to perform the activities within this practice. For example, the robot is expected to know how to make coffee and read the newspaper.
    \end{itemize}
In \cite{dignumFSPT2017} there is a first formalization of all these aspects based on dynamic logic. Due to space limitations we will not include this whole formalization here, but just discuss a few points that are important for the current implementation of social practices in Jason.\\
The core element of the social practice for an agent is the plan pattern, which gives it handles to plan its behaviour. Plan patterns are defined as follows:
\begin{definition}{{\bf Plan Patterns Language}}\\
{\rm A Plan Pattern of a social practice is an element of the set PP, which is the smallest set closed under}
\begin{gather*}
\gamma\phi \in PP\\
\gamma_1\phi_1,\gamma_2\phi_2 \in PP \Rightarrow \gamma_1\phi_1 +\gamma_2\phi_2 \in PP \\
\gamma_1\phi_1,\gamma_2\phi_2 \in PP \Rightarrow \gamma_1\phi_1 \&\gamma_2\phi_2 \in PP \\
\gamma_1\phi_1,\gamma_2\phi_2 \in PP \Rightarrow \gamma_1\phi_1 ;\gamma_2\phi_2 \in PP\\
\end{gather*}
\end{definition}
Here $\gamma\phi$ stands for any sequence of actions $\gamma$ that contain actions contributing towards the achievement of $\phi$ (starting from a particular situation). $\phi$ is the \emph{purpose} of that part of the practice. There can be more effects, but they are not all specified. So, in our morning routine practice the plan pattern can be defined as:\\
$\gamma_1\phi_1;(\gamma_2\phi_2\&\gamma_3\phi_3);\gamma_4\phi_4$\\
$\phi_1$ is the person is awake\\
$\phi_2$ is the coffee is served and $\phi_3$ is the pills are taken\\
$\phi_4$ is the person is mentally stimulated.\\
So, the purpose of the first part of the morning routine is that the person is awake. This might be done by opening the curtains, giving some loud noise or otherwise. If the purpose is achieved by opening the curtains, not only is the person awake, but the curtains are also open. The latter is merely a side effect of achieving the purpose in this way. 

Two more things
should be noted
about these patterns. One is that the overall pattern is supposed to achieve the overall purpose of the social practice. This is a formal constraint, but we only treat this implicitly. The other
is that after a part of the plan pattern is finished, it automatically triggers the start of the next part of the pattern. In the full formalism this is 
assured, but is not explicit from only this fragment. In the same way, a social practice is started when the start condition becomes true. 
It then
becomes available for execution and can be used by any agent present in the situation.

Finally, the formalism of social practices also guarantees that there is a common belief in the elements of the social practice and if actions are taken everyone has the at least a common belief about the effects in as far as they are important for the social practice. Thus it guarantees a common situation awareness.

\section{The care robot scenario}
\label{sec:scenario}

In this section we elaborate on the care robot scenario outlined in the introduction, and describe how we have modelled and implemented it using Jason.

We assume the high-level operation of the robot is based on a BDI interpreter, and that it comes equipped with goals and plans to trigger and enact its care activities (most likely with some customisation of key parameters possible). In this section we consider only a small subset of the robot's duties: to wake the patient at a certain time in the morning, to provide coffee as required, and to provide mental stimulation. We do not specify any goals of the robot outside the practice here, but normally the care robot would also have its own goals such as powering its battery, (vacuum) cleaning a room and taking care of the health of the patient. The morning routine can be seen as a part of the plan to take care of the health of the patient, while there might be no social practice for vacuum cleaning and this is completely handled by the standard BDI part of the care robot.

Social practices provide patterns of coordination for multiple agents in terms of landmark states rather than explicit sequences of actions. Therefore they do not make limiting assumptions about the temporal aspects of actions and their effects leading up to a landmark. Only the landmarks themselves are explicitly temporally ordered. To illustrate this we include some temporal complexity in the scenario by including durative actions (i.e.~those that take place across an interval of time), an action with a delayed effect, and a joint durative action, which has its desired effect only if two participants perform it during overlapping time intervals. Durative and joint actions are implemented using a Jason metainterpreter\footnote{A metainterpreter is a programming language interpreter written in the same, or a similar, language. It can be used to prototype extensions to the base language.} that is described in Section~\ref{sec:implementation}. To simulate the passing of time, we use a ``ticker'' agent with a recursive plan that periodically performs a tick action to update the time recorded in the environment. We use Jason's synchronous execution mode, so the robot, patient and ticker agents perform a single reasoning step in every step of the simulation.

Figure~\ref{fig:robot_plans} shows the robot's initial beliefs, rules and plans. It has four sets of plans (lines 25 onwards). These have declarative goals (i.e.~their triggering goals express desired states) and use Jason preprocessing directives to transform them according to a predefined declarative achievement goal pattern~\cite{JasonBook}.

\begin{figure}[tb]
  \newlength{\codewidth}
  \setlength{\codewidth}{\textwidth}
  \addtolength{\codewidth}{-4ex}
\hbox{
\rule{1.5ex}{0pt}
\begin{minipage}{\codewidth}
\begin{Verbatim}
// Initial beliefs and rules

durative(makePodCoffee).
durative(readNewspaper).
joint(readNewspaper).
durative_action_continuation_pred
    (readNewspaper, continueReadingNewspaper).
durative_action_continuation_pred(makePodCoffee,
                                  continueMakingPodCoffee).
continueReadingNewspaper :-
    started(readNewspaper, T1) &
    not started_durative_action(readNewspaper, patient, _) &
    time(T2) & 
    T2 <= T1 + 20.
continueReadingNewspaper :-
    started_durative_action(readNewspaper, patient, _) &
    not stopped_durative_action(readNewspaper, patient, _).

continueMakingPodCoffee :- state(coffee, not_made).

wake_up_phrase("Good morning sleepyhead!").

// Plans

{begin ebdg(state(patient,awake))}
+!state(patient,awake) : wake_up_phrase(P) <- talkToPatient(P).
+!state(patient,awake) <- shakePatient.
+!state(patient,awake) <- 
    openCurtains;
    .wait(state(patient, awake), 30000).
{end}

{begin ebdg(state(patient,mentally_stimulated))}
+!state(patient, mentally_stimulated) <-
    .wait(state(patient, awake)); .fail.
+!state(patient, mentally_stimulated) <- play_mozart.
+!state(patient, mentally_stimulated) <- 
    !solve([body_term(
              "", readNewspaper[participants([patient,robot])])]).
{end}

+!state(coffee, served) <-
    !state(coffee, made);
    serveCoffee.

{begin ebdg(state(coffee,made))}
+!state(coffee, made) :
        resource(coffee_pods) & resource(coffee_pod_machine) <-
    makePodCoffee;
    .wait(state(coffee, made), 10000).
+!state(coffee, made) : resource(instant_coffee) <-
    makeInstantCoffee.
{end}

{ include("metainterpreter.asl") }
\end{Verbatim}
\end{minipage}
}
  \caption{Plans for the care robot domain}
  \label{fig:robot_plans}
\end{figure}

The first set of plans (lines 25--31) are for achieving a state where the patient is awake, with alternative plans for talking to the patient, shaking him, and opening the curtains and waiting for the light to wake him. The \emph{exclusive backtracking declarative goal} (``edbg'') pattern specifies that additional failure-handling logic should be added to ensure that all the plans will be tried (once each) until the goal is achieved, or all plans fail. The action of opening the curtains has a delayed effect: it will eventually wake the patient\footnote{Actions are implemented in Jason by defining an \emph{execute} method in Java class modelling the environment. The delay is currently hard-coded in this class.}.

The second set of plans (lines 33--40) are used for the goal of having the patient mentally simulated, and also use the ebdg pattern. The first plan waits for the patient to be awake, and then fails so that the other plans will be tried. The other two alternatives involve playing the music of Mozart to the patient, and initiating the joint action of reading the newspaper with the patient. As joint actions are not directly supported by Jason, lines 38--39 call this action via the \verb|solve| goal that is handled by our metainterpreter.

These plans are followed by a single plan for serving coffee. This has the subgoal of having the coffee made, and then the action of serving the coffee is performed.

The final set of plans are for reaching a state in which the coffee is made. The options are to use a coffee pod and wait for it to finish (up to a time limit)\footnote{The wait timeouts (in ms) in these plans are for simulation purposes only, and would be much longer in a real-world application.}, or to make instant coffee.

The initial segment of the listing contains initial beliefs and rules related to the processing of durative actions: declarations of which actions are declarative and/or joint, and predicates and associated rules defining the circumstances in which the robot will continue performing the durative actions.

The environment sends a percept to all participants of a joint action when any other participant performs the action for the first time or performs a stop action with the joint action as an argument. The patient agent has a plan to take his pills once he is awake. He also has a plan that will respond to the robot beginning the joint newspaper reading action by also beginning that action. He will continue reading the newspaper for 40 time units if he is in a good mood, but only 20 if it is in a bad mood. Being woken by daylight (after the curtains are opened) leaves him in a good mood; being shaken awake leaves him in a bad mood, and talking will not wake him up. Thus, if the robot begins with goals to have the patient awake and mentally stimulated, the patient will be left in a bad mood by being shaken awake and the newspaper reading will be shorter (and less stimulating) that if he were in a good mood.

\section{A care robot with social practices}
\label{sec:scenario_revisited}

Section~\ref{sec:scenario} introduced the care robot scenario. In this section, we consider how the robot could be enhanced using social practices.

As noted previously, it is assumed that the robot comes equipped with appropriate goals and plans, and that it is possible to customise certain parameters such as the time the user likes to wake up, and the time and style of coffee that he likes to have. However, customising each plan in isolation will not easily provide the coordination between activities and dynamic adaptability to different contexts that can be provided by social practices. To perform most effectively, the robot should choose, for a given context, the plans for each goal that will achieve the best outcomes for the patient, and furthermore, consider constraints on goal orderings that arise from preferences and habit. For example, if the patient prefers to be woken at a certain time in a given context (e.g.~when his family is due to visit) and/or in a certain way (e.g.~by the curtains being opened), his mood is likely to be adversely affected if he is woken at a different time, and his engagement with subsequent activities (such as reading the newspaper together) may be reduced. In this section we describe how this type of contextual information can be addressed by the use of a social practice.
%

In Section~\ref{sec:scenario}, we described the various plans and actions available to the robot. We now assume that the following ``morning routine'' social practice has emerged\footnote{It is beyond the scope of this paper to consider how social practices might be learned and/or communicated.}. We present this as a set of beliefs in the form used by our social practice reasoning plans that will be discussed in Section~\ref{sec:implementation}. Note that we only illustrate a small subset of what would be likely to be a real morning routine for a patient and his/her care robot, but this is sufficient to highlight the nature of social practices and their relation to BDI agents.

\begin{Verbatim}
social_practice(morningRoutine,
    [state(location, home), resource(coffee_pods),
     resource(coffee_pod_machine), resource(pills),
     resource(newspaper_subscription),
     (time(T) & T < 1200)]).
   
landmark(morningRoutine, pa,
     [], [action(robot, openCurtains)],
     state(patient, awake)).

landmark(morningRoutine, pt,
    [pa], [action(patient, takePills)],
    state(pills, taken)).

landmark(morningRoutine, cs,
    [pa], [action(robot, makePodCoffee)],
    state(coffee, served)).
    
landmark(morningRoutine, ms,
    [pt,cs], [action([robot,patient], readNewspaper)],
    state(patient, mentally_stimulated)).
\end{Verbatim}

The first belief above encodes the name of the social practice and a list of conditions that must all hold for it to become active: there are constraints on the location, the resources available, and the time (here, the number 1200 is a proxy for some real-world time that ends the morning routine period).

The other four beliefs model the landmarks, specifying the social practice they are part of, an identifier for the landmark, a list of landmarks that must have been reached previously, a list of actions and their actors that are associated with the landmark, and finally, a goal that is the purpose of the landmark. The landmarks are: (1) to have the patient awake due to the robot opening the curtains, (2) for the patient to have taken his pills, (3) to have the coffee served, which should involve the robot making pod coffee, and (4) for the patient to be mentally stimulated due to the newspaper being read jointly. These landmarks are partially ordered with 1 before 2 and 3, which both precede 4.

Comparing this social practice to the robot plans shown in Figure~\ref{fig:robot_plans}, it can be seen that it avoids an ineffective attempt to wake the patient by talking to him, and prevents him from being left in a bad mood after being shaken awake. It agrees with the first-ordered plan for making coffee (by making pod coffee), and avoids an ill-fated attempt by the robot to provide mental stimulation by playing Mozart. Furthermore, it specifies an ordering on these activities that is not intrinsic to the plans themselves. Note also, that the social practice does not provide complete information on how to reach the landmark of having coffee served: it indicates that the robot should make pod coffee, but doesn't specify the action of serving the coffee. While a planning system could deduce the missing action using a model of actions and their effects~\cite{Miller}, a BDI agent does not have this capability. Instead, a BDI agent using social practices must reason about how its existing plans could be used to satisfy landmarks given potentially incomplete information about the actions it must perform.

Furthermore, the robot may already have goals to wake the patient, provide mental stimulation, etc., and the activation of a social practice should not create additional instances of those goals. Thus, the activation of a social practice should override the agent's normal behaviour (for the relevant goals) during the period of activation.

As social practices are structured in terms of ordered landmarks, which model expected states to be reached in a pattern of inter-agent coordination, it is necessary for the agent to actively monitor the status of landmarks once their prior landmarks have been achieved, and to actively work towards the fulfilment of the current landmarks for which it has associated actions. In the next section, we present a meta-deliberation cycle for Jason agents that addresses this and the other issues outlined above, and which enables the successful execution of our care robot enhanced with social practices.

\section{Implementation}
\label{sec:implementation}

\subsection{Meta-level reasoning about social practices}
\label{sec:metadeliberation}

\begin{figure}[tb]
\hbox{
\rule{1.5ex}{0pt}
\begin{minipage}{\codewidth}
\begin{Verbatim}
/* Rules */
// Omitted: has_plan_generating_action/4 and for_all/1
relevant_sp(SP) :-
    social_practice(SP, Requirements) & forall(Requirements).
sp_selection(Options, CurrentSP) :-
    selected_sp(CurrentSP) & .member(CurrentSP, Options).
sp_selection([SP|_], SP).

/* Initial goal */
!metaDeliberate.

/* Plans */
@metaplan[atomic]
+!metaDeliberate <-
    .findall(SP, ( relevant_sp(SP) & not completed_sp(SP) ),
             RelevantSPs);
    if (RelevantSPs == []) {
        if (selected_sp(CurrentlySelectedSP)) {
            -selected_sp(CurrentlySelectedSP)    
        }
    } else {
        if ( sp_selection(RelevantSPs, SelectedSP) &
	     not selected_sp(SelectedSP) ) {
           -+selected_sp(SelectedSP)
        }
        for (monitored(Purpose, SP, ID)) {
            if (Purpose) { +completed_landmark(SP, ID, Purpose) } 
        }
    }
    .wait(500);
    !!metaDeliberate.
    
+selected_sp(SP) <-
    for (landmark(SP, ID, _, _, Purpose)) {
        PurposeNoAnnots[dummy] = Purpose[dummy];
        if (.intend(PurposeNoAnnots)) {
             .suspend(PurposeNoAnnots);
             +suspended_intention(SP, ID, PurposeNoAnnots)
        }
        .add_plan({@suspend_purpose(SP,ID)
                   +!PurposeNoAnnots <- .suspend(PurposeNoAnnots)},
                  landmark(SP,ID), begin)
    }
    for (landmark(SP, ID, [], Actions, Purpose)) {
        !activate_landmark(SP, ID, Actions, Purpose)
    }.
    
@activate_landmark[atomic]
+!activate_landmark(SP, ID, Actions, Purpose) <- 
    PurposeNoAnnots[dummy] = Purpose[dummy];
    +monitored(PurposeNoAnnots, SP, ID)
    if (Actions = [action(Actors, Act)] &
          (Actors = Me | (.list(Actors) & .member(Me, Actors)))) {
        if (has_plan_generating_action(
	        {+!Purpose}, Act, BodyTerms, Path)) {
            !!solve([body_term("!",Purpose)], Path)   
        } else {
        if (joint(Act)) {
            !!solve([body_term("", Act[participants(Actors)])])
        } else {
        if (durative(Act)) {
            !!solve([body_term("", Act)])
        } else {
        if (Act =.. [F, _, _] & .substring(".", F)) {
            !!solve([body_term(".", Act)])
        } else {
	    Act
        }}}}
    } else { .print("Multiple actions are not yet supported"); }.
    
@completed_landmark[atomic]
+completed_landmark(SP, ID, Purpose) <-
    .succeed_goal(Purpose);
    -monitored(Purpose, SP, ID);
    .remove_plan(suspend_purpose(SP,ID));
    for ( landmark(SP, ID2, PrecedingLMs, Actions, Purpose2) &
          not completed_landmark(SP, ID2, _) &
          .findall(PrecID, (.member(PrecID, PrecedingLMs) &
                            completed_landmark(SP, PrecID, _)),
                   CompletedPrecIDs) &
          .difference(PrecedingLMs, CompletedPrecIDs, []) ) {
      !activate_landmark(SP, ID2, Actions, Purpose2)
    }
    .findall(ID2, ( landmark(SP, ID2, _, _,_) &
                    not completed_landmark(SP, ID2, _) ),
             PendingLandmarks);
    if (PendingLandmarks == []) { +completed_sp(SP) }.
\end{Verbatim}
\end{minipage}
}
  \caption{Rules and plans for social practice reasoning}
  \label{fig:metadeliberation}
\end{figure}

Maintaining awareness of social practices (SPs), and contributing to them in an appropriate way, requires agents to detect when each known social practice becomes active or inactive, to monitor the state of the landmarks in an active social practice, and to trigger the appropriate activity if an active SP has an action for the agent associated with the next landmark. This is a type of meta-level reasoning that the agent should perform periodically, and it may override the performance of any standard BDI processing of goals, which is not informed by social practices. We note that, on an abstract level, the same was done in \cite{Gentile} where the plan pattern was translated into a global pattern in Drools (Java based expert system) and the specific interactions within each phase were programmed in a chatbot.

The question then arises of how best to implement such a meta-level reasoner in a BDI architecture. While the best performance can, no doubt, be achieved by extending a BDI platform using its underlying implementation language, this approach requires significant knowledge of the implementation and requires using an imperative coding style that is not best suited to reasoning about goals~\cite{DBLP:journals/ijaose/Logan18} and for rapid prototyping and dissemination of new reasoning techniques. Therefore, in this work we define the meta-level reasoner as a plan for a \verb|metadeliberate| goal that reasons about social practices, sleeps and then calls itself recursively. This, and some other plans it triggers, are shown in Figure~\ref{fig:metadeliberation}. The plans make use of some extensions to Jason, handled by a metainterpreter that is described in the following subsection\footnote{See \url{https://github.com/scranefield/jason-social-practices} for source code.}.

First, we give a brief overview of the syntax of the AgentSpeak syntax, as implemented (and extended) in Jason. Based on logic-programming languages like Prolog, the basic language constructs are atoms (beginning with lower case letters), variables (beginning with upper case letters), and structured terms with a functor and terms as arguments. Jason also allows terms to have \emph{annotations}: lists of terms in square brackets, and these are treated specially during unification\footnote{\url{https://github.com/jason-lang/jason/blob/master/doc/tech/annotations.adoc}}. Agent programs consist of initial beliefs, rules (Horn clauses) and plans. Plans are expressed using the syntax \verb|@|\emph{Label} \emph{Trigger} \verb|:| \emph{ContextCond} \verb|<-| \emph{PlanBody}, where \emph{Label}, \emph{ContextCond} and \emph{PlanBody} are optional. Triggers are events such as the creation of a new goal (\verb|+!|\emph{GoalTerm}) or a new belief (\verb|+|\emph{BeliefTerm}). \emph{ContextCond} is a logical formula stating when the plan is applicable, and is evaluated using beliefs and rules. A plan body contains a sequence of actions (terms with no prefix), built-in internal actions (terms whose functor contains a ``.''), belief additions and deletions (with prefix `+' and `-', respectively), queries over beliefs and rules (with prefix `?'), and subgoals (with prefix `!' or `!!'---the latter creates a separate intention for the subgoal). Some control structures such as if-then-else are also supported within a plan body.


\hyphenation{meta-Deliberate}

\begin{sloppy}
  The social practice reasoner runs in response to the goal \verb|meta|-\verb|Deliberate| (line 10 in Figure~\ref{fig:metadeliberation}). Lines 13 to 31 show the plan for this goal. The \verb|atomic| annotation on the plan label ensures that steps of this plan are not interleaved with steps of other plans. The plan begins by (re)considering which social practice (if any) should be active. It uses the rules in lines 3 to 7 to find social practices that are relevant (i.e.~all their requirements hold), and to select one (currently, the first option is always selected). If none are relevant (lines 17--20), any existing belief about the currently selected social practice is retracted. Otherwise (lines 22--28), if the selection has changed, the belief about the selection is updated. Any monitored landmarks are then checked to see if their purpose has been fulfilled (lines 26--28). If so, a belief about their completion is added. The plan then sleeps for period, before triggering itself to be re-run in a new intention (lines 30--31). The new intention is needed for the recursive call as the plan is atomic, and the agent's other plans must be allowed to run).
\end{sloppy}

A new belief about a selected social practice is handled by the plan in lines 33--46. This loops through the landmarks to check if the agent already has intentions to achieve any of their purposes\footnote{The unifications in lines 35 and 50 instantiate the variable on the left with the value of the variable on the right, but with any annotations removed.}. If so, these intentions are suspended, and this is recorded in a belief so the intentions can be later marked as successful if the landmark is completed (see line 73). A plan is also temporarily added (lines 40-42) to ensure that if some other active plan of the agent separately creates this intention, it will be immediately suspended (the new plan is placed before any existing plans for that goal). For each landmark in the social practice that has no prior landmarks, a goal is created to activate it (lines 44--46).

Landmark activations are handled by the plan in lines 48--69. A belief recording that the landmark's purpose should be monitored is added, then the action associated with the landmark is processed (only a single action is supported currently). If the action is to be performed by the agent, three options are considered. First (line 54), a query is made to find a solution for achieving the landmark's purpose that involves performing the specified action. A set of rules (not shown) handle this query by searching for the action recursively (up to a prespecified depth bound) through the plans that achieve the purpose, and the subgoals in those plans, and so on. The plans' context conditions are checked for the top level plans (those for the landmark's purpose), but the recursive calls do not, as, in general, it cannot be known how the state of the world will change as these plans are executed. If such a solution is found, it is recorded as a goal-plan tree ``path'' (see Section~\ref{sec:metainterpreter}) and passed to a call to our Jason metainterpreter (line 56). If no such solution is found, and the action is a joint, durative or internal one, the metainterpreter is called to handle this (lines 58--65. Otherwise, the action is performed directly (line 67).

Finally, the plan in lines 71--87 handles completed landmarks---those for which the purpose has been achieved. Any suspended intentions for the purpose are succeeded, the belief stating that the landmark should be monitored is retracted, and the temporary plan added in lines 40-42 is removed. The plan then checks for subsequent landmarks that should now be activated (if all their prior landmarks are completed), and finally adds a belief that the social practice has completed if all its landmarks are completed. Another plan (not shown) handles social practices that become inactive when their relevance conditions cease to hold. In this case, any active landmarks should be abandoned, and original intentions to achieve their purposes can be resumed.

\todo{Elaborate next sentence if time?}
With these plans and the metadeliberation goal in place, our robot and patient agent can successfully coordinate their actions across the landmarks of the social practice, ensuring that the patient remains in a good mood, and engages in the newspaper reading for longer.

\subsection{A Jason metainterpreter}
\label{sec:metainterpreter}

\begin{figure}[tb]
\hbox{
\rule{1.5ex}{0pt}
\begin{minipage}{\codewidth}
\begin{Verbatim}
/* Rules */
context_ok(plan(_,_,ContextCond,_)) :- ContextCond.

// Rules for filter_list/3 omitted

/* Plans */

// Entry points: solve body term list, with or without a
// prespecified path through the goal-plan tree

+!solve(PlanBodyTerms) <- !solve(PlanBodyTerms, [], 1, no_path).
+!solve(PlanBodyTerms, Path) <- !solve(PlanBodyTerms, [], 1, Path).

// solve body term list

+!solve([], _, _, _).
+!solve([body_term(Prefix, Term)|BTs], Intn, N, Path) <-
    Intn2 = [N|Intn];
    !solve(Prefix, Term, Intn2, Path);
    !solve(BTs, Intn, N+1, Path).

// solve body terms

+!solve("?", B, _, _) <- ?B.
+!solve("+", B, _, _) <- +B.
+!solve("-", B, _, _) <- -B.
+!solve("!", solve(PBTs), Intn, Path) <-
    !solve(PBTs, Intn, 1, Path).
+!solve("!", G, Intn, Path) <-
    meta.relevant_plan_bodies_as_terms({+!G}, RPlans);
    if (.list(Path) & Path = [N|PathTail] &
        .nth(N, RPlans, plan(Label,_,_,PlanBodyTerms))) {
        !solve(PlanBodyTerms, [Label|Intn], 1, PathTail);
    } else {
        ?filter_list(RPlans, context_ok, APlans);
        !solve_one(APlans, Intn, Path);
    }.
+!solve("", AnnotatedAction, _, _) :
        Act[dummy] = AnnotatedAction[dummy] & durative(Act) <-
    ?durative_action_continuation_pred(Act, Query);
    if (durative_action_cleanup_goal(Act, CleanupGoal)) {
        CUGoal = CleanupGoal;
    } else {
        CUGoal = true;
    }
    if (joint(Act) & Act[participants(P)] = AnnotatedAction) {
        ParticipantAnnotation = [participants(P)];
    } else {
        ParticipantAnnotation = [];
    }
    if (time(T)) {
        Act[durative|ParticipantAnnotation];
        -+started(Act, T)[source(meta)];
    }
    !solve_durative(Query, Act, ParticipantAnnotation, CUGoal).
+!solve("", Action, _, _) <- Action.
+!solve(".", .fail, _, _) <- .fail.
+!solve(".", .print(S), _, _) <- .print(S).
+!solve(".", .wait(Cond), _, _) <- .wait(Cond).
+!solve(".", .wait(Cond, Timeout), _, _) <- .wait(Cond, Timeout).

// Solve some plan in a list of plans

+!solve_one([plan(Label,_,_,PlanBodyTerms)|_], Intn, Path)
        : not tried_plan(Label, Intn) <-
    +tried_plan(Label, Intn);
    !solve(PlanBodyTerms, [Label|Intn], 1, Path);
    -tried_plan(Label, Intn).
        
-!solve_one([_|PlanTerms], Intn, Path) <-
    !solve_one(PlanTerms, Intn, Path).

+!solve_durative(Query, Act,
                 ParticipantAnnotation, CleanupGoal) <-
    if (Query) {
        Act[durative|ParticipantAnnotation];
        !solve_durative(Query, Act,
	                ParticipantAnnotation, CleanupGoal);
    } else {
        stop(Act)[durative|ParticipantAnnotation];
        if (CleanupGoal \== true) {
            !CleanupGoal;
        }
    }.
-!solve_durative(_, Act, _) <-
    -started(Act, _).
\end{Verbatim}
\end{minipage}
}
  \caption{A Jason metainterpreter}
  \label{fig:metainterpreter}
\end{figure}

Figure~\ref{fig:metainterpreter} shows our Jason metainterpreter, which extends the AgentSpeak metainterpreter defined by Winikoff~\cite{DBLP:conf/promas/Winikoff05}, and specialises it for use with Jason. The metainterpreter is initiated by calling a \verb|solve| goal with a list of plan body terms, i.e.~terms representing the various types of goals and actions that can appear in a plan body. In each \verb|plan_body| term, the \verb|Prefix| argument identifies the type of the goal or action by a string (e.g.~`?' for a query to the belief base, `+' for a belief addition, and `!' for a subgoal). From line 16 onwards, each \verb|solve| trigger event has additional arguments that: (a) identify the current intention as a stack of current subgoal indices within each active plan body, interleaved with the labels for the plans currently active to solve those subgoals, and (b) a final \verb|Path| argument, explained below. The intention identifier is used in lines 64 to 71, which sequentially try the plans for a goal, asserting beliefs about the plans that have been tried. Lines 70 and 71 leverage Jason's failure-handling mechanisms (posting achievement goal deletion events upon goal failure) to detect that an attempt to ``solve'' a plan failed, and to try the next plan. Finally, note that there are two work-arounds for current restrictions of Jason. First, as Jason does not provide a way to decompose a plan body from within plans, line 30 calls a custom internal action we have implemented in Java. Given a trigger event (e.g.~a new goal event), this action returns a list of relevant plans, encoded as list of plan terms, each including a list of \verb|plan_body| terms. Second, internal (in-built) actions cannot be called dynamically via instantiated higher order variables (as used for other actions: see line 56). Therefore, lines 57 to 60 enumerate specific internal actions that are supported (and more can be added).

We made the following extensions to support new capabilities;
\begin{enumerate}
\item
  Durative actions, as required by our scenario, are supported (lines 38--55 and 73--86)\footnote{The first context condition on line 39 instantiates variable \texttt{Act} to the action term, with any Jason annotations removed.}. A continuation predicate, and optionally a clean-up goal\footnote{Cleanup goals can be used to remove any temporary state recorded as beliefs during a durative action's execution, but are not important for the discussion in this paper.}, for the action are looked up (lines 40--45), the time the action was started is recorded as a belief (line 53), and a \verb|solve_durative| goal is created (line 55) to trigger the performance of the action. The plan for this goal (lines 73--84) checks the continuation condition (passed as variable \verb|Query|). It is intended that the query is a 0-arity predicate defined by a rule in the agent's program. If the query succeeds, the action is executed with a ``durative'' annotation (which the environment should check for), and possibly an annotation listing the action participants if it is a joint action (see below). The goal is then called recursively. If the query fails, \verb|stop(Act)| is executed (again, with the appropriate annotations). Thus, durative actions are implemented by repeated execution of an action until the corresponding stop action is called.
\item
  Joint actions are also supported. These are durative actions with an annotation listing the intended action participants. The environment should notify all intended participants (via a percept) when a durative action is called for the first time or is stopped, thus enabling the participants to coordinate their actions. It should also keep a history of the time intervals over which the participants perform the action, as its outcome will depend on the existence and length of a period of overlap.
\item
  As explained in Section~\ref{sec:metadeliberation}, when a landmark in a social practice includes an action associated with the current agent, the plan to activate a landmark attempts to find an existing plan that can achieve the landmark's purpose while also including the specified action. This is a recursive search through plans and their subgoals, and it results in a pre-selected path through the goal-plan tree~\cite{DBLP:conf/atal/LoganTY17} corresponding to the search space for satisfying the landmark's purpose. This path can be passed to the metainterpreter (line 12), to guide it directly to the pre-chosen subplans, and eventually the desired action. This feature is useful for plan pre-selection in other meta-reasoning contexts as well, e.g.~choosing plans based on their effect on the values of a human user~\cite{DBLP:conf/ijcai/CranefieldWDD17}.
\end{enumerate}


\section{Conclusions}
We have argued that for interactive settings, as sketched in our scenario, the use of social practices is a good compromise between using a fixed interaction protocol and deliberation and planning from scratch at each point during the interaction. We proposed a mechanism for a BDI agent to maintain awareness about and contribute towards the completion of social practices, and presented this as a meta-deliberation plan for Jason agents. To extend Jason with features required for this plan and our care robot scenario, we also presented a Jason metainterpreter. These contributions can serve as a specification of potential extensions to the BDI reasoning cycle, but also allow the approach to be directly applied within Jason agents.


Our approach allows BDI agents to use their existing plans to achieve social practice landmarks that do not detail all actions required to achieve the landmark. However, there are some subtleties that remain to be explored. For example, suppose that an agent has a plan for a goal that is the purpose of a landmark, but that one of that plan's subgoals is the purpose of a prior landmark. In that case, the execution of the plan should be adapted to exclude this subgoal. In future work we intend to investigate more complex cases such as this. We also intend to develop elaborate scenarios that use all aspects of a social practice, and compare these with agent implementations where no social practice is used, both in terms of the outcomes of the agent and the ease of design of the agents.




\clearpage

\bibliographystyle{ACM-Reference-Format}  
\bibliography{aamas2019_social_practices}  


\begin{thebibliography}{00}


\ifx \showCODEN    \undefined \def \showCODEN     #1{\unskip}     \fi
\ifx \showDOI      \undefined \def \showDOI       #1{#1}\fi
\ifx \showISBNx    \undefined \def \showISBNx     #1{\unskip}     \fi
\ifx \showISBNxiii \undefined \def \showISBNxiii  #1{\unskip}     \fi
\ifx \showISSN     \undefined \def \showISSN      #1{\unskip}     \fi
\ifx \showLCCN     \undefined \def \showLCCN      #1{\unskip}     \fi
\ifx \shownote     \undefined \def \shownote      #1{#1}          \fi
\ifx \showarticletitle \undefined \def \showarticletitle #1{#1}   \fi
\ifx \showURL      \undefined \def \showURL       {\relax}        \fi
\providecommand\bibfield[2]{#2}
\providecommand\bibinfo[2]{#2}
\providecommand\natexlab[1]{#1}
\providecommand\showeprint[2][]{arXiv:#2}

\bibitem[\protect\citeauthoryear{Augello, Gentile, and Dignum}{Augello
  et~al\mbox{.}}{2016}]%
        {Gentile}
\bibfield{author}{\bibinfo{person}{Agnese Augello}, \bibinfo{person}{Manuel
  Gentile}, {and} \bibinfo{person}{Frank Dignum}.}
  \bibinfo{year}{2016}\natexlab{}.
\newblock \showarticletitle{Social Practices for Social Driven Conversations in
  Serious Games}. In \bibinfo{booktitle}{{\em GALA 2015}},
  \bibfield{editor}{\bibinfo{person}{Alessandro De~Gloria} {and}
  \bibinfo{person}{Remco Veltkamp}} (Eds.). \bibinfo{publisher}{Springer
  International Publishing}, \bibinfo{pages}{100--110}.
\newblock


\bibitem[\protect\citeauthoryear{Bordini, H{\"u}bner, and Wooldridge}{Bordini
  et~al\mbox{.}}{2007}]%
        {JasonBook}
\bibfield{author}{\bibinfo{person}{Rafael~H. Bordini},
  \bibinfo{person}{Jomi~Fred H{\"u}bner}, {and} \bibinfo{person}{Michael
  Wooldridge}.} \bibinfo{year}{2007}\natexlab{}.
\newblock \bibinfo{booktitle}{{\em Programming multi-agent systems in
  {AgentSpeak} using {Jason}}}.
\newblock \bibinfo{publisher}{Wiley}.
\newblock


\bibitem[\protect\citeauthoryear{{Bourdieu (trans. R. Nice)}}{{Bourdieu (trans.
  R. Nice)}}{1972}]%
        {Bourdieu1976}
\bibfield{author}{\bibinfo{person}{P. {Bourdieu (trans. R. Nice)}}.}
  \bibinfo{year}{1972}\natexlab{}.
\newblock \bibinfo{booktitle}{{\em Outline of a theory of practice}}.
\newblock \bibinfo{publisher}{Cambridge University Press}.
\newblock


\bibitem[\protect\citeauthoryear{Cranefield, Winikoff, Dignum, and
  Dignum}{Cranefield et~al\mbox{.}}{2017}]%
        {DBLP:conf/ijcai/CranefieldWDD17}
\bibfield{author}{\bibinfo{person}{Stephen Cranefield},
  \bibinfo{person}{Michael Winikoff}, \bibinfo{person}{Virginia Dignum}, {and}
  \bibinfo{person}{Frank Dignum}.} \bibinfo{year}{2017}\natexlab{}.
\newblock \showarticletitle{No Pizza for You: Value-based Plan Selection in
  {BDI} Agents}. In \bibinfo{booktitle}{{\em Proceedings of the Twenty-Sixth
  International Joint Conference on Artificial Intelligence}}.
  \bibinfo{publisher}{ijcai.org}, \bibinfo{pages}{178--184}.
\newblock


\bibitem[\protect\citeauthoryear{Dignum}{Dignum}{2018}]%
        {dignumFSPT2017}
\bibfield{author}{\bibinfo{person}{F. Dignum}.}
  \bibinfo{year}{2018}\natexlab{}.
\newblock \showarticletitle{Interactions as Social Practices: towards a
  formalization}.
\newblock \bibinfo{journal}{{\em arXiv\/}} (\bibinfo{year}{2018}).
\newblock
\showURL{%
\url{https://arxiv.org/abs/1809.08751}}


\bibitem[\protect\citeauthoryear{Dignum and Dignum}{Dignum and Dignum}{2015}]%
        {COIN2014}
\bibfield{author}{\bibinfo{person}{V. Dignum} {and} \bibinfo{person}{F.
  Dignum}.} \bibinfo{year}{2015}\natexlab{}.
\newblock \showarticletitle{Contextualized Planning Using Social Practices}. In
  \bibinfo{booktitle}{{\em Coordination, Organizations, Institutions and Norms
  in Agent Systems X: COIN 2014}},
  \bibfield{editor}{\bibinfo{person}{A.~Ghose}, \bibinfo{person}{N.~Oren},
  \bibinfo{person}{P.~Telang}, {and} \bibinfo{person}{Thangarajah J}} (Eds.).
  \bibinfo{publisher}{Springer-Verlag}, \bibinfo{pages}{36--52}.
\newblock


\bibitem[\protect\citeauthoryear{Folsom-Kovarik and Schatz}{Folsom-Kovarik and
  Schatz}{2014}]%
        {Flosom2014}
\bibfield{author}{\bibinfo{person}{J.T. Folsom-Kovarik} {and}
  \bibinfo{person}{S. Schatz}.} \bibinfo{year}{2014}\natexlab{}.
\newblock \showarticletitle{{AI} challenge problem: Scalable models for
  patterns of life}.
\newblock \bibinfo{journal}{{\em AI Magazine\/}} \bibinfo{volume}{35},
  \bibinfo{number}{1} (\bibinfo{year}{2014}), \bibinfo{pages}{10--14}.
\newblock


\bibitem[\protect\citeauthoryear{Giddens}{Giddens}{1979}]%
        {Giddens1979}
\bibfield{author}{\bibinfo{person}{A. Giddens}.}
  \bibinfo{year}{1979}\natexlab{}.
\newblock \bibinfo{booktitle}{{\em Central problems in social theory: Action,
  structure and contradiction in social analysis}}.
\newblock \bibinfo{publisher}{University of California Press}.
\newblock


\bibitem[\protect\citeauthoryear{Harel, Yumak, and Dignum}{Harel
  et~al\mbox{.}}{2018}]%
        {Harel}
\bibfield{author}{\bibinfo{person}{Raoul Harel}, \bibinfo{person}{Zerrin
  Yumak}, {and} \bibinfo{person}{Frank Dignum}.}
  \bibinfo{year}{2018}\natexlab{}.
\newblock \showarticletitle{Towards a Generic Framework for Multi-party
  Dialogue with Virtual Humans}. In \bibinfo{booktitle}{{\em Proceedings of the
  31st International Conference on Computer Animation and Social Agents}} {\em
  (\bibinfo{series}{CASA 2018})}. \bibinfo{publisher}{ACM},
  \bibinfo{address}{New York, NY, USA}, \bibinfo{pages}{1--6}.
\newblock


\bibitem[\protect\citeauthoryear{Logan}{Logan}{2018}]%
        {DBLP:journals/ijaose/Logan18}
\bibfield{author}{\bibinfo{person}{Brian Logan}.}
  \bibinfo{year}{2018}\natexlab{}.
\newblock \showarticletitle{An agent programming manifesto}.
\newblock \bibinfo{journal}{{\em International Journal of Agent-Oriented
  Software Engineering\/}} \bibinfo{volume}{6}, \bibinfo{number}{2}
  (\bibinfo{year}{2018}), \bibinfo{pages}{187--210}.
\newblock


\bibitem[\protect\citeauthoryear{Logan, Thangarajah, and Yorke-Smith}{Logan
  et~al\mbox{.}}{2017}]%
        {DBLP:conf/atal/LoganTY17}
\bibfield{author}{\bibinfo{person}{Brian Logan}, \bibinfo{person}{John
  Thangarajah}, {and} \bibinfo{person}{Neil Yorke-Smith}.}
  \bibinfo{year}{2017}\natexlab{}.
\newblock \showarticletitle{Progressing Intention Progression: {A} Call for a
  Goal-Plan Tree Contest}. In \bibinfo{booktitle}{{\em Proceedings of the 16th
  Conference on Autonomous Agents and MultiAgent Systems, {AAMAS} 2017,
  S{\~{a}}o Paulo, Brazil, May 8-12, 2017}}. \bibinfo{publisher}{{ACM}},
  \bibinfo{pages}{768--772}.
\newblock


\bibitem[\protect\citeauthoryear{Mercuur, Dignum, and Kashima}{Mercuur
  et~al\mbox{.}}{2017}]%
        {Mercuur}
\bibfield{author}{\bibinfo{person}{Rijk Mercuur}, \bibinfo{person}{Frank
  Dignum}, {and} \bibinfo{person}{Yoshihisa Kashima}.}
  \bibinfo{year}{2017}\natexlab{}.
\newblock \showarticletitle{Changing {Habits} {Using} {Contextualized}
  {Decision} {Making}}. In \bibinfo{booktitle}{{\em Advances in {Social}
  {Simulation} 2015}}, \bibfield{editor}{\bibinfo{person}{Wander Jager},
  \bibinfo{person}{Rineke Verbrugge}, \bibinfo{person}{Andreas Flache},
  \bibinfo{person}{Gert de~Roo}, \bibinfo{person}{Lex Hoogduin}, {and}
  \bibinfo{person}{Charlotte Hemelrijk}} (Eds.). \bibinfo{publisher}{Springer
  International Publishing}, \bibinfo{pages}{267--272}.
\newblock


\bibitem[\protect\citeauthoryear{Miller, Dignum, and Dignum}{Miller
  et~al\mbox{.}}{2018}]%
        {Miller}
\bibfield{author}{\bibinfo{person}{Tim Miller}, \bibinfo{person}{Virginia
  Dignum}, {and} \bibinfo{person}{Frank Dignum}.}
  \bibinfo{year}{2018}\natexlab{}.
\newblock \showarticletitle{Planning for Human-Agent collaboration using Social
  Practices}. In \bibinfo{booktitle}{{\em First international workshop on
  socio-cognitive systems at IJCAI 2018}}.
\newblock


\bibitem[\protect\citeauthoryear{Minsky}{Minsky}{1988}]%
        {Minsky}
\bibfield{author}{\bibinfo{person}{Marvin Minsky}.}
  \bibinfo{year}{1988}\natexlab{}.
\newblock \showarticletitle{A framework for representing knowledge}.
\newblock In \bibinfo{booktitle}{{\em Readings in Cognitive Science}},
  \bibfield{editor}{\bibinfo{person}{Allan Smith} {and} \bibinfo{person}{Edward
  Collins}} (Eds.). \bibinfo{publisher}{Morgan Kaufmann},
  \bibinfo{pages}{156--189}.
\newblock


\bibitem[\protect\citeauthoryear{Narasimhan, Roberts, Xenitidou, and
  Gilbert}{Narasimhan et~al\mbox{.}}{2017}]%
        {Gilbert}
\bibfield{author}{\bibinfo{person}{Kavin Narasimhan}, \bibinfo{person}{Thomas
  Roberts}, \bibinfo{person}{Maria Xenitidou}, {and} \bibinfo{person}{Nigel
  Gilbert}.} \bibinfo{year}{2017}\natexlab{}.
\newblock \showarticletitle{Using ABM to Clarify and Refine Social Practice
  Theory}. In \bibinfo{booktitle}{{\em Advances in Social Simulation 2015}},
  \bibfield{editor}{\bibinfo{person}{Wander Jager}, \bibinfo{person}{Rineke
  Verbrugge}, \bibinfo{person}{Andreas Flache}, \bibinfo{person}{Gert de~Roo},
  \bibinfo{person}{Lex Hoogduin}, {and} \bibinfo{person}{Charlotte Hemelrijk}}
  (Eds.). \bibinfo{publisher}{Springer International Publishing},
  \bibinfo{pages}{307--319}.
\newblock


\bibitem[\protect\citeauthoryear{Reckwitz}{Reckwitz}{2002}]%
        {Reckwitz}
\bibfield{author}{\bibinfo{person}{A. Reckwitz}.}
  \bibinfo{year}{2002}\natexlab{}.
\newblock \showarticletitle{Toward a Theory of Social Practices}.
\newblock \bibinfo{journal}{{\em European Journal of Social Theory\/}}
  \bibinfo{volume}{5}, \bibinfo{number}{2} (\bibinfo{year}{2002}),
  \bibinfo{pages}{243--263}.
\newblock


\bibitem[\protect\citeauthoryear{Schatzki}{Schatzki}{2012}]%
        {Schatzki2012}
\bibfield{author}{\bibinfo{person}{Theodore~R. Schatzki}.}
  \bibinfo{year}{2012}\natexlab{}.
\newblock \bibinfo{booktitle}{{\em A Primer on Practices}}.
\newblock \bibinfo{publisher}{SensePublishers}, \bibinfo{address}{Rotterdam},
  \bibinfo{pages}{13--26}.
\newblock


\bibitem[\protect\citeauthoryear{Shove, Pantzar, and Watson}{Shove
  et~al\mbox{.}}{2012}]%
        {shove2012}
\bibfield{author}{\bibinfo{person}{E. Shove}, \bibinfo{person}{M. Pantzar},
  {and} \bibinfo{person}{M. Watson}.} \bibinfo{year}{2012}\natexlab{}.
\newblock \bibinfo{booktitle}{{\em The Dynamics of Social Practice}}.
\newblock \bibinfo{publisher}{Sage}.
\newblock


\bibitem[\protect\citeauthoryear{Winikoff}{Winikoff}{2006}]%
        {DBLP:conf/promas/Winikoff05}
\bibfield{author}{\bibinfo{person}{Michael Winikoff}.}
  \bibinfo{year}{2006}\natexlab{}.
\newblock \showarticletitle{An {AgentSpeak} Meta-interpreter and Its
  Applications}.
\newblock In \bibinfo{booktitle}{{\em Programming Multi-Agent Systems, Third
  International Workshop, ProMAS 2005}}. \bibinfo{series}{Lecture Notes in
  Computer Science}, Vol.~\bibinfo{volume}{3862}.
  \bibinfo{publisher}{Springer}, \bibinfo{pages}{123--138}.
\newblock


\end{thebibliography}

\end{document}